\documentclass[%
 jmp,
 amsmath,
 amssymb,
 reprint
]{revtex4-1}

\allowdisplaybreaks

\usepackage[bottom]{footmisc}
\usepackage[hidelinks]{hyperref}
\usepackage{graphicx}
\usepackage{dcolumn}
\usepackage{bm}

\newcommand{\edth}{\textnormal{\dh}}
\newcommand{\thorn}{\textnormal{\th}}

\usepackage[utf8]{inputenc}
\usepackage[T1]{fontenc}
\usepackage{mathptmx}

\usepackage{float}
\floatstyle{plain}

\begin{document}


\title{
   A symmetric hyperbolic formulation of the vacuum Einstein equations
   in affine-null coordinates 
}

\author{Justin L. Ripley}
 \email{lloydripley@gmail.com}
\affiliation{ 
DAMTP,
Centre for Mathematical Sciences,
University of Cambridge,
Wilberforce Road, Cambridge CB3 0WA, UK.
}%

\date{\today}

\begin{abstract}
   We present a symmetric hyperbolic formulation of the Einstein
   equations in affine-null coordinates.
   Giannakopoulos et. al. (T. Giannakopoulos, 
   D. Hilditch, and M. Zilhao, Phys. Rev. D
   102, 064035 (2020), arXiv:2007.06419 [gr-qc])
   recently showed that the most commonly numerically implemented
   formulations of the Einstein equations in affine-null coordinates
   (and other single-null coordinate systems)
   are only weakly--but not strongly--hyperbolic.
   By making use of the tetrad-based Newman-Penrose formalism,
   our formulation avoids the hyperbolicity problems of
   the formulations investigated by Giannakopoulos et. al.
   We discuss a potential application of our formulation for studying
   gravitational wave scattering.
\end{abstract}

\maketitle
\section{\label{sec:introduction}Introduction}

Bondi-like coordinate systems have played an important
role in understanding gravitational radiation
and the structure of future null infinity
in asymptotically flat spacetimes\cite{Bondi:1960jsa,
doi:10.1098/rspa.1962.0206,doi:10.1098/rspa.1962.0161}.
One of the main attractions of using outgoing Bondi-like
coordinates is that the
Einstein equations near future null infinity take a relatively
simple form when one foliates with an outgoing light-sheet.
Moreover, the equations of motion have a nested structure,
which can be convenient for analyzing the structure of the equations,
and numerically solving them.
In numerical relativity, outgoing Bondi-like coordinates
have found use in \emph{Cauchy-characteristic extraction}
schemes to accurately extract outgoing gravitational waves
near future null infinity
\cite{Tamburino:1966zz,Bishop:1996gt,Handmer:2016mls,Winicour:2008vpn}.
A specific variant of Bondi-like coordinates,
ingoing affine-null coordinates, have found extensive use
in numerical studies of black hole/black branes in asymptotically
anti de Sitter (AdS) spacetimes (for a review, see \cite{Chesler:2013lia}).
Ingoing affine-null coordinates have been useful in those
contexts as they simplify the study of gravitational
radiation near black hole/black brane horizons.

Despite their longtime use, the hyperbolicity properties of the Einstein
equations in Bondi-like coordinate systems was only recently
studied by Giannakopoulos et. al. \cite{Giannakopoulos:2020dih}.
Surprisingly, those authors found that the Einstein
Equations, when written as an evolution system for the metric
in those Bondi and affine-null coordinates, formed only
a weakly--but not strongly--hyperbolic system of partial
differential equations for the metric components.
Following the general approach of 
\cite{doi:10.1098/rspa.1981.0045,Luk:2011vf,Hilditch:2019maz},
we show that it is relatively
straightforward to write down a symmetric
(and hence strongly) hyperbolic formulation
of the Einstein equations when one works in the 
tetrad-based based Newman-Penrose \cite{Newman_Penrose_paper}
formalism\footnote{We note that the possibility of a strongly hyperbolic
formulation of the Einstein equations for Bondi-like
coordinate systems existing for a tetrad-based formalism was suggested
by Giannakopoulos et. al. \cite{Giannakopoulos:2020dih}.}.
Unfortunately, our symmetric hyperbolic evolution system does not have 
the same convenient nested structure as the Einstein
equations written as an evolution system for the metric components
in Bondi-like coordinates.
Moreover, working in the Newman-Penrose formalism significantly
increases the number of dynamical variables one must solve for.
Despite these potential drawbacks,
the formulation could have some applications
in studying gravitational wave scattering, which we discuss
in Secs.~\ref{sec:Initial data} and \ref{sec:discussion}. 

Strong hyperbolicity is a
necessary condition for system of partial differential equations (PDE)
to have a well posed initial value problem for generic initial data
\cite{kreiss1989initial}.
Moreover, numerical solutions to a system of PDE can
only converge to the continuum
solution if the system is well-posed\footnote{As is stressed
in \cite{Giannakopoulos:2020dih} though,
this does not imply that a numerical solution
of a weakly hyperbolic system will necessarily exhibit any pathological
behavior at a given \emph{fixed} resolution}.
As not all formulations of the Einstein equations are strongly
hyperbolic, any given formulation of the equations must be
explicitly checked
(for further discussion and review
see, e.g. \cite{Sarbach:2012pr,Hilditch:2013sba}).
We review the concepts of weak, strong, and symmetric hyperbolicity
in Appendix~\ref{sec:hyperbolicity_review}.

Compared to work on Bondi-like (or single-null) coordinate
systems, there is an extensive literature
on the hyperbolicity of the Einstein equations
in \emph{double null coordinates}, which we briefly review.
The first major result goes back to
Friedrich\cite{doi:10.1098/rspa.1981.0045}, who showed
that the Einstein equations form a strongly hyperbolic system
in a tetrad formulation when using double null coordinates.
Friedrich further proved a local existence of solutions near the
intersection of the two initial characteristic data surfaces.
Luk later provided an improved existence result
(following earlier work by Rendall\cite{doi:10.1098/rspa.1990.0009})
using a tetrad formalism\cite{Luk:2011vf}; which was 
later reformulated in the Newman-Penrose formalism by
Hilditch et. al.\cite{Hilditch:2019maz}.
While we believe that it is likely
that our formulation has a well-posed initial value
problem, we are unaware of any theorem that demonstrates
that strongly hyperbolic single-null formulations of 
the Einstein equations have a well-posed initial
value problem, and it is beyond the scope of this article
to provide such a theorem here\footnote{For a review which
discusses earlier attempts
to formulate a well-posed initial value problem
for single-null formulations of the Einstein equations,
see \cite{Winicour:2008vpn}.}.

Our sign conventions for the metric, Riemann tensor, etc
follow Newman and Penrose\cite{Newman_Penrose_paper}
(see also Chandrasekhar\cite{Chandrasekhar_bh_book});
e.g. we use $+---$ signature for the metric.
We work in four spacetime dimensions.
We index spacetime indices $(0,1,2,3)$ with lower-case Greek letters, and
index angular indices $(2,3)$ with lower-case Latin letters.
An overbar represent complex conjugation.
\section{\label{sec:single_null_coordinates}
   Choice of tetrad for affine-null coordinates
   }
\label{sec:coordinates_tetrad_choice}
We will use the same tetrad choice as is used in \cite{Newman_Penrose_paper}\S6.
We first write down the metric for outgoing affine-null coordinates
(we consider the case of ingoing affine-null coordinates
in Appendix~\ref{sec:ingoing_formalism}):
\begin{align}
\label{eq:outgoing_affine_null}
   ds^2
   =&
      Vdu^2
   +  2dudr
   \nonumber\\&
   -  h_{ab}\left(d\theta^a+W^adu\right)\left(d\theta^b+W^bdu\right)
   .
\end{align}
We have used our four coordinate degrees of freedom to
set $g_{rr}=g_{ra}=0$, and $g_{ru}=1$.
With these conditions, we see that $u$ is a null coordinate:
($g^{\alpha\beta}\partial_{\alpha}u\partial_{\beta}u=0$).
We see with this choice of coordinates that
$r$ is an affine parameter for outgoing null geodesics.

As in \cite{Newman_Penrose_paper}\S6, we choose the following tetrad:
\begin{subequations}
\label{eq:outgoing_null_tetrad}
\begin{align}
\label{eq:outgoing_l_tetrad}
   l^{\mu}\partial_{\mu}
   =&
      \partial_r
   ,\\
\label{eq:outgoing_n_tetrad}
   n^{\mu}\partial_{\mu}
   =&
      \partial_u
   +  P\partial_r
   +  R^a\partial_a
   ,\\
\label{eq:outgoing_m_tetrad}
   m^{\mu}\partial_{\mu}
   =&
      Q\partial_r
   +  S^a\partial_a
   .
\end{align}
\end{subequations}
where $Q$ and $S^a$ are complex.
Relating the tetrad to the metric via the identity
$
   g^{\mu\nu}
   =
   l^{\mu}n^{\nu}
   +
   n^{\mu}l^{\nu}
   -
   m^{\mu}\bar{m}^{\nu}
   -
   \bar{m}^{\mu}m^{\nu}
   ,
$
(see Appendix \ref{sec:review_NP_formalism}), we find that
\begin{subequations}
\begin{align}
   g^{rr}
   =
   -  V
   =&
      2\left(P - Q\bar{Q}\right)
   ,\\
   g^{ra}
   =
   -  W^a
   =&
      R^a
   -  \bar{S}^aQ
   -  S^a\bar{Q}
   ,\\
   g^{ab}
   =
   -  h^{ab}
   =&
   -  S^a\bar{S}^b
   -  S^b\bar{S}^a
   .
\end{align}
\end{subequations}
The tetrad we have chosen satisfies
$\kappa=\mathcal{R}\epsilon=0$.
There are three remaining tetrad degrees of freedom,
which can be expressed as:
\begin{align}
   n^{\mu}
   \to
   n^{\mu}
   ,\qquad
   l^{\mu}
   \to
      \bar{z}m^{\mu}
   +  z\bar{m}^{\mu}
   +  z\bar{z}n^{\mu}
   ,\qquad
   m^{\mu}
   \to
      m^{\mu}
   +  zn^{\mu}
   ,\nonumber\\
   n^{\mu}
   \to
   n^{\mu}
   ,\qquad
   l^{\mu}
   \to
   l^{\mu}
   ,\qquad
   m^{\mu}
   \to
   e^{i x}m^{\mu}
   ,\nonumber
\end{align}
where $z$ is complex and $x$ is real, which we use to set
$\pi=0$ and $\mathcal{I}\epsilon=0$, respectively.
To summarize, we have used our six tetrad rotations to set
\begin{align}
   \label{eq:tetrad_conditions}
   \kappa=\pi=\epsilon=0
   . 
\end{align}
More physically, we have used our six tetrad degrees of freedom to
make $l^{\mu}$ geodesic, and to parallel propagate
$n^{\mu}$ and $m^{\mu}$ along that geodesic:
$
   l^{\mu}\nabla_{\mu}l^{\nu}
   =
   l^{\mu}\nabla_{\mu}n^{\nu}
   =
   l^{\mu}\nabla_{\mu}m^{\nu}
   =0
$
\section{Symmetric hyperbolic formulation of the Einstein equations}
\label{sec:outgoing_eom}
In this section we present our symmetric hyperbolic formulation
of the Einstein equations, using our coordinates and tetrad
from Sec.~\ref{sec:coordinates_tetrad_choice}.
To simplify some of our expressions, we will make use the
Geroch-Held-Penrose (GHP)
derivative operators\cite{GHP_paper} $\edth$, $\edth'$ and $\thorn'$.
As we have set $\epsilon=0$, the operators $\thorn$ and $D$
are equivalent. We will continue to use $D$ to emphasize
the special role the outgoing null vector $l^{\mu}$ plays in
this formulation of the Einstein equations.

We obtain evolution equations for the
coefficients of $l^{\mu}$ and $m^{\mu}$
by using the tetrad commutation relations
\eqref{eq:commutation_relations}:
\begin{subequations}
\label{eq:constraint_equations_tetrad}
\begin{align}
   D n^{\mu}
   +
   \left(\gamma+\bar{\gamma}\right)l^{\mu}
   -
   \bar{\tau} m^{\mu}
   -
   \tau\bar{m}^{\mu}
   =&
   0
   ,\\
   D m^{\mu}
   +
   \left(\bar{\alpha}+\beta\right)l^{\mu}
   -
   \bar{\rho} m^{\mu}
   -
   \sigma\bar{m}^{\mu}
   =&
   0
   .
\end{align}
\end{subequations}
Eqs.~\eqref{eq:constraint_equations_tetrad}
should be read as expressions for the
individual components of $l^{\mu}$ and $m^{\mu}$.
Writing things out in terms of components, we have
\begin{subequations}
\begin{align}
   DP 
   + \left(\gamma+\bar{\gamma}\right)
   -  \bar{\tau}Q
   -  \tau\bar{Q}
   =&
   0,\\
   D R^a
   -  \bar{\tau}S^a
   -  \tau\bar{S}^a
   =&
   0,\\
   DQ
   +  \left(\bar{\alpha}+\beta\right)
   -  \bar{\rho}Q
   -  \rho\bar{Q}
   =&
   0,\\
   DS^a
   -  \bar{\rho}S^a
   -  \rho\bar{S}^a
   =&
   0
   .
\end{align}
\end{subequations}
From the Ricci identities, Eqs.~\eqref{eq:riem-all}, we can write down
the following system for the nonzero Ricci rotation coefficients:
\begin{subequations}
\label{eq:constraint_equations_ricci_rotation}
\begin{align}
   D\rho-\rho^2+\sigma\bar{\sigma}
   =&
   0
   ,\\
   D\sigma-\left(\rho+\bar{\rho}\right)\sigma-\Psi_0
   =&
   0
   ,\\
   D\tau-\rho\tau-\sigma\bar{\tau}-\Psi_1
   =&
   0
   ,\\
   D\alpha-\rho\alpha-\bar{\sigma}\beta
   =&
   0
   ,\\
   D\beta-\sigma\alpha-\bar{\rho}\beta-\Psi_1
   =&
   0
   ,\\
   D\gamma-\tau\alpha-\bar{\tau}\beta-\Psi_2
   =&
   0
   ,\\
   D\lambda-\rho\lambda-\bar{\sigma}\mu
   =&
   0
   ,\\
   D\mu-\bar{\rho}\mu-\sigma\lambda-\Psi_2
   =&
   0
   ,\\
   D\nu-\bar{\tau}\mu-\tau\lambda-\Psi_3
   =&
   0
   .
\end{align}
\end{subequations}
We can rewrite the Bianchi identities to obtain the following
evolution system for the Weyl scalars:
\begin{subequations}
\label{eq:evo_weyl_scalars}
\begin{align}
   \label{eq:evo_Psi_0}
   \left(\thorn'+\mu\right)\Psi_0
   -
   \left(\edth-4\tau\right)\Psi_1
   -
   3\sigma\Psi_2
   &=
   0
   ,\\
   \left(D+\thorn'+2\mu-4\rho\right)\Psi_1
   -
   \left(\edth-3\tau\right)\Psi_2
   \nonumber\\
   -
   \left(\edth'+\nu\right)\Psi_0
   -2\sigma\Psi_3
   &=
   0
   ,\\
   \left(D+\thorn'+3\mu-3\rho\right)\Psi_2
   -
   \left(\edth-2\tau\right)\Psi_3
   &\nonumber\\
   -
   \left(\edth'+2\nu\right)\Psi_1
   +
   \lambda\Psi_0
   -
   \sigma\Psi_4
   &=
   0
   ,\\
   \left(D+\thorn'+4\mu-2\rho\right)\Psi_3
   -
   \left(\edth-\tau\right)\Psi_4
   &\nonumber\\
   -
   \left(\edth'+3\nu\right)\Psi_3
   +
   2\lambda\Psi_1
   &=
   0
   ,\\
   \left(D-\rho\right)\Psi_4
   -
   \edth'\Psi_3
   +
   3\lambda\Psi_2
   &=
   0
   .
\end{align}
\end{subequations}
We obtained this system by adding together
(\eqref{eq:bianchi-1}+\eqref{eq:bianchi-6}),
(\eqref{eq:bianchi-2}+\eqref{eq:bianchi-7}),
and
(\eqref{eq:bianchi-3}+\eqref{eq:bianchi-8}).
With this, we see that we have specified a system of evolution/constraint
equations for all of the nonzero Newman-Penrose scalars
and metric components (via the tetrad components of $n^{\mu}$ and $m^{\mu}$).

To determine the hyperbolicity of the evolution
system defined by Eqs.
\eqref{eq:constraint_equations_tetrad},
\eqref{eq:constraint_equations_ricci_rotation}, 
and
\eqref{eq:evo_weyl_scalars},
we write down the principal part:
\begin{subequations}
\begin{align}
   \begin{pmatrix}
      \mathcal{D}_1^{\mu} 
      & 
      0 
      &
      0
      \\
      0 
      & 
      \mathcal{D}_2^{\mu}
      & 
      0 
      \\
      0
      &
      0
      &
      \mathcal{D}_3^{\mu}
   \end{pmatrix}
   \partial_{\mu}
   \begin{pmatrix}
      \vec{\Psi}
      \\
      \vec{\Gamma}
      \\
      \vec{g}
   \end{pmatrix}
   ,
\end{align}
\end{subequations}
where
\begin{subequations}
\begin{align}
   \mathcal{D}_1^{\mu}
   \equiv&
   \begin{pmatrix}
      n^{\mu} & -m^{\mu} & 0 & 0 & 0
      \\
      -\bar{m}^{\mu} & l^{\mu}+n^{\mu} & -m^{\mu} & 0 & 0
      \\
      0 & -\bar{m}^{\mu} & l^{\mu}+n^{\mu} & -m^{\mu} & 0 
      \\
      0 & 0 & -\bar{m}^{\mu} & l^{\mu}+n^{\mu} & -m^{\mu}
      \\
      0 & 0 & 0 & -\bar{m}^{\mu} & l^{\mu} 
   \end{pmatrix}
   ,\\
   \mathcal{D}_2^{\mu}
   \equiv&
   id_9\times l^{\mu}
   ,\\
   \mathcal{D}_3^{\mu}
   \equiv&
   id_8\times l^{\mu}
   ,\\
   \vec{\Psi}^T
   \equiv&
   \left(
      \Psi_0
      ,
      \Psi_1
      , 
      \Psi_2
      ,
      \Psi_3
      ,
      \Psi_4
   \right)
   ,\\
   \vec{\Gamma}^T
   \equiv&
   \left(
      \rho,\sigma,\tau,\alpha,\beta,\gamma,\lambda,\mu,\nu
   \right)
   ,\\
   \vec{g}^T
   \equiv&
   \left(n^v,n^r,n^a,m^v,m^r,m^a\right)
   ,
\end{align}
\end{subequations}
($id_n$ denotes the $n\times n$ identity matrix).
By inspection, we see that the principal symbol
\begin{align}
   \begin{pmatrix}
      \mathcal{D}_1^{\mu} 
      & 
      0 
      &
      0
      \\
      0 
      & 
      \mathcal{D}_2^{\mu}
      & 
      0 
      \\
      0
      &
      0
      &
      \mathcal{D}_3^{\mu}
   \end{pmatrix}
   \xi_{\mu}
   ,
\end{align}
is Hermitian. 
Moreover, we see that it is positive definite with 
respect to the timelike direction 
$t^{\mu}\equiv l^{\mu}+n^{\mu}$:
\begin{align}
   t_{\mu}
   \begin{pmatrix}
      \mathcal{D}_1^{\mu} 
      & 
      0 
      &
      0
      \\
      0 
      & 
      \mathcal{D}_2^{\mu}
      & 
      0 
      \\
      0
      &
      0
      &
      \mathcal{D}_3^{\mu}
   \end{pmatrix}
   =
   \mathrm{diag}\left(
      1,2,2,2,1,d_9,d_8
   \right)
   ,
\end{align}
where $d_n\equiv\mathrm{diag}\left(id_n\right)$.
We conclude that the evolution system defined by
defined by Eqs,~\eqref{eq:evo_weyl_scalars},
\eqref{eq:constraint_equations_ricci_rotation}, and
\eqref{eq:constraint_equations_tetrad}
is symmetric hyperbolic with respect to the timelike direction $t^{\mu}$
(c.f. \cite{Hilditch:2019maz} for a similar recent construction, 
but which instead works in double null coordinates).
\section{\label{sec:Initial data}
   Incoming gravitational wave initial data
}
Given that $l^{\mu}\partial_{\mu}=\partial_r$, we see that
Eqs.~\eqref{eq:constraint_equations_ricci_rotation},
\eqref{eq:constraint_equations_tetrad} form a system of ``constraint''
equations for the Ricci rotation coefficients and tetrad components
along a $u=const.$ hypersurface.
Using the Bianchi identities
\eqref{eq:bianchi-1}-\eqref{eq:bianchi-4},
we can write down a system of constraint
equations for the Weyl scalars $\Psi_1,\Psi_2,\Psi_3$, and $\Psi_4$:
\begin{subequations}
\label{eq:constraint_weyl_scalars}
\begin{align}
   \left(D-4\rho\right)\Psi_1
   -
   \edth'\Psi_0
   &=
   0
   ,\\
   \left(D-3\rho\right)\Psi_2
   -
   \edth'\Psi_1
   +
   \lambda\Psi_0
   &=
   0
   ,\\
   \left(D-2\rho\right)\Psi_3
   -
   \edth'\Psi_2
   +
   2\lambda\Psi_1
   &=
   0
   ,\\
   \left(D-\rho\right)\Psi_4
   -
   \edth'\Psi_3
   +
   3\lambda\Psi_1
   &=
   0
   .
\end{align}
\end{subequations}
Given these equations, provided we have constraint satisfying
initial data at $r=r_{min}$ on our $u=u_i$ initial data
surface, the only free initial data left to specify are the
real and imaginary components of $\Psi_0$.
The remaining technical challenge with this initial data setup is
to find consistent initial data for
the constrained variables at the point $r=r_{min}$ of
our initial data surface.
One simple prescription is to have all of the Newman-Penrose
scalars satisfy a known, exact solution to the Einstein equations
at $r=r_{min}$: $\alpha=\alpha^{(bkgrd)}, \Psi_4=\Psi_4^{(bkgrd)}$, etc.
Then, over a \emph{compact} region (in $r$)
on the $u=u_i$ initial data surface, one sets
\begin{align}
   \Psi_0 = {}^{(bkgrd)}\Psi_0 + {}^+\Psi_0
   ,
\end{align}
where ${}^{(bkgrd)}\Psi_0$ is the known, ``background'' solution to $\Psi_0$,
and ${}^+\Psi_0$ is the free initial specification of $\Psi_0$.
We can think of ${}^+\Psi_0$ as specifying an incoming gravitational
wave added on top of a given exact solution to the Einstein equations.
Given the exact background solution, we have consistent initial
data at $r=r_{min}$, and we can then use the constraint equations
Eqs.
\eqref{eq:constraint_equations_ricci_rotation},
\eqref{eq:constraint_equations_tetrad},
and
\eqref{eq:constraint_weyl_scalars}
to specify consistent
initial data over the entire $u=const.$ hypersurface.
As we evolve in time, we must continue to specify consistent
boundary data at $r=r_{min}$.
For all time before the ``packet'' of ${}^+\Psi_0$
reaches that boundary, we can continue to set
all of the Newman-Penrose scalars to their background values
at that boundary.
After that point, one would need to specify evolution
equations for the other Newman-Penrose scalars
on the boundary to continue to have consistent boundary
data there.

\begin{figure}
\centering
\includegraphics[width=\linewidth]{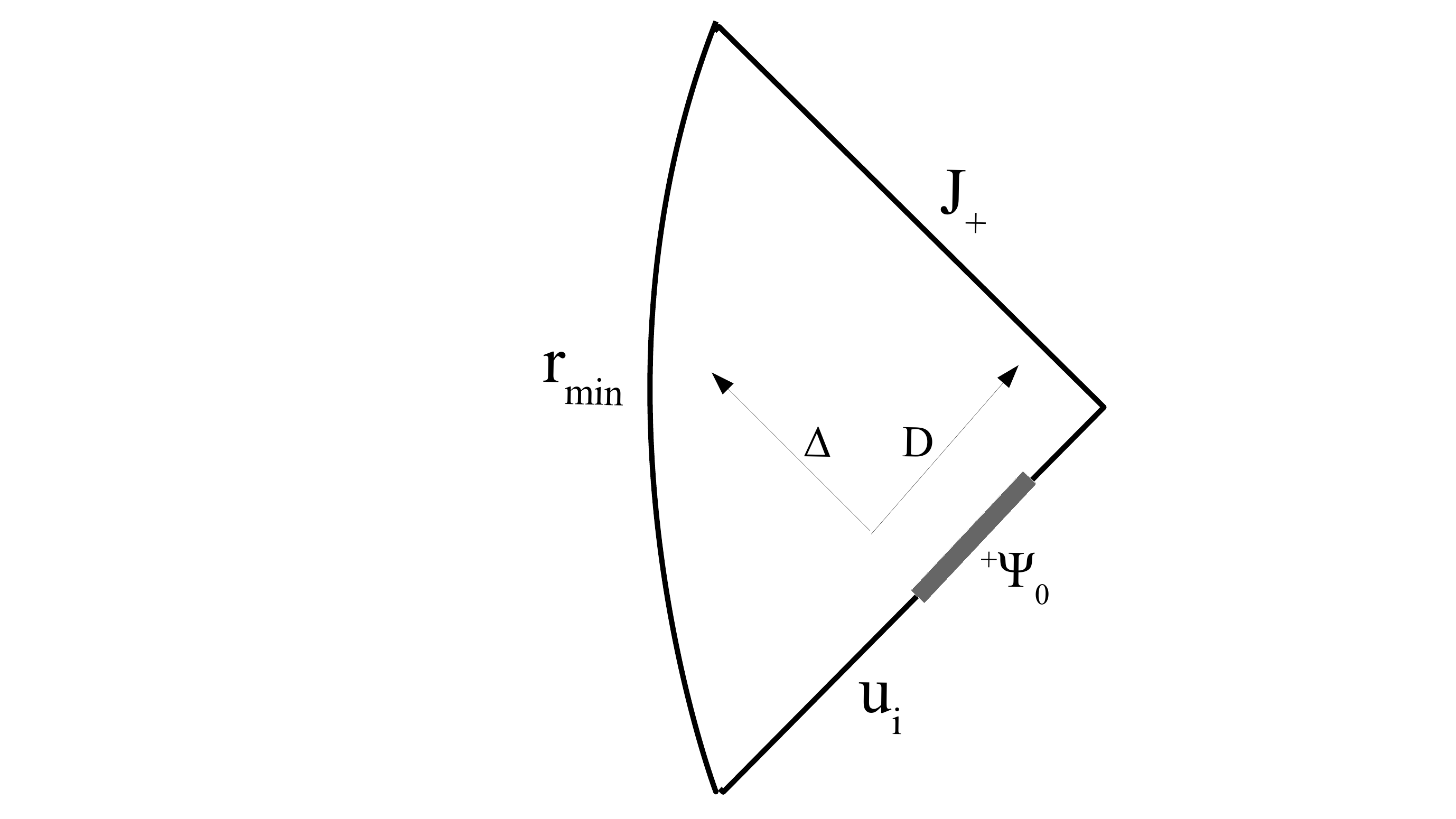}
\caption{Schematic Penrose diagram for purely ingoing gravitational
   radiation initial data.
   On the initial $u=u_i$ initial data surface, one specifies
   ${}^+\Psi_0$ over a compact region.
   All of the other Ricci rotation coefficients, Weyl scalars, and
   metric components can then be found on the initial
   data surface by integrating their
   respective evolution/constraint equations. 
   At the boundary $r=r_{min}$ all Newman-Penrose scalars are set to
   be their ``background'' values; this is consistent initial data
   for the times before the
   ``packet'' of ${}^+\Psi_0$ reaches that boundary.
   See Sec.~\ref{sec:Initial data} for more discussion.
   }
\label{fig:integration_diagram}
\end{figure}

We conclude that in this particular
setup we have specified purely ingoing gravitational
wave initial data, with no outgoing gravitational waves at
the boundary $r=r_{min}$ (provided the background solution is stationary).
See Fig.~\ref{fig:integration_diagram} for a schematic illustration.

\section{\label{sec:discussion}
   Discussion
}
While single-null coordinates provide several conceptual and technical
advantages over Cauchy-like evolution schemes, the most
commonly used formulations of the Einstein equations in those
coordinate systems were recently shown to be only weakly
--but not strongly--hyperbolic \cite{Giannakopoulos:2020dih}.
We have shown that by using the
\emph{tetrad}-based Newman-Penrose formalism,
it is relatively straightforward to construct
a symmetric hyperbolic formulation of the Einstein equations
in affine-null coordinates, which are example of a single null
(or Bondi-like) coordinate system.
We have also presented a method to construct purely ingoing gravitational
wave initial data, which could be used to analytically and/or numerically
study gravitational wave scattering problems.
While we have only considered outgoing affine-null coordinates
\eqref{eq:outgoing_affine_null}, our formulation can be straightforwardly
adapted to ingoing affine-null coordinates
by applying the GHP ``prime'' operator\cite{GHP_paper}
to all of our equations; see Appendix~\ref{sec:ingoing_formalism}
for more details.

As strong hyperbolicity is a necessary property
for the well-posedness of the initial value formulation
of a PDE, the condition is widely regarded as a crucial
property a system of PDE must have in
order for that system to be amenable to stable numerical evolution,
(see, e.g. \cite{Sarbach:2012pr,Hilditch:2013sba}).
This being said,
numerical evolution of weakly hyperbolic systems may only give
non-convergent/pathological results with initial data that is sufficiently
``rough'', and at sufficiently high numerical resolution
\cite{Giannakopoulos:2020dih}.
The stable, convergent evolution of many numerical schemes that have used
single-null (and weakly hyperbolic) formulations of the Einstein equations
could potentially be explained by the fact that those earlier studies
worked with smooth initial data,
and use numerical dissipation (which is typically
required for stable numerical evolution of even strongly hyperbolic
systems, see, e.g. \cite{gustafsson1995time}).

While we have derived a strongly hyperbolic formulation of the
Einstein equations, we have not proven that that system
has a well posed initial value problem.
While Rendall \cite{doi:10.1098/rspa.1990.0009} and 
Luk \cite{Luk:2011vf} have provided general theorems that demonstrate
local existence of solutions for strongly hyperbolic formulations
of double null formulations of the Einstein equations, we work
with a single-null formulation, and thus their results are not
directly applicable to our formulation.
We leave to future work a suitable  generalization of their results
to single-null strongly hyperbolic systems. 

We caution that from a practical standpoint we have not considered
other important qualities a system of PDE
must satisfy in order to admit stable numerical evolution.
For example, we have not analyzed the stability of
``constraint violating modes'' in the evolution system,
whose potential  growth away from the constraint surface
could effectively rule out practical any numerical evolution
of the system (for more discussion, see, e.g. 
\cite{doi:10.1063/1.532694,Gundlach:2005eh}).
More generally, we have not computed any bound on the region for which a 
solution to the system we have presented would have well-posed evolution
off of the initial data hypersurface.
Studying the domain of existence of solutions could be studied using
the techniques used by the authors of, e.g. \cite{Hilditch:2019maz}.

In this article we have only provided a prescription for initial data
that involves an ingoing gravitational wave
(although in Appendix \ref{sec:ingoing_formalism}
we discuss how one could set up outgoing
gravitational wave initial data when working in
ingoing affine-null coordinates).
We restricted ourselves to that kind of initial data due to the
form of the evolution/constraint
Eqs.
\eqref{eq:constraint_equations_tetrad},
\eqref{eq:constraint_equations_ricci_rotation},
and
\eqref{eq:constraint_weyl_scalars}: with that system we could
use the real and imaginary parts of
$\Psi_0$ as free initial data, provided we had constraint
satisfying initial data at the boundary $r=r_{min}$ of our
initial data ($u=const.$) surface.
The most straightforward way of satisfying this
constraint is by setting the boundary to be a known solution to the
Einstein equations. A more detailed analysis of the 
Newman-Penrose equations may reveal a better/more generic
way of providing constraint equations
at $r=r_{min}$\footnote{For example,
this is done in \cite{Hilditch:2019maz}.
In their case it is more straightforward to do this as
with their tetrad condition the $\delta$ derivatives only involve
``angular'' derivatives, so the Newman-Penrose equations
\eqref{eq:riem-11}-\eqref{eq:riem-13}
provide a convenient set of constraint equation at the
$r=r_{min}$}.
While we leave constructing more sophisticated initial data sets in this
formulation to future work, we note that already the initial data 
setup we provide could allow for investigations of gravitational wave
scattering to future null infinity (for the outgoing null coordinates
considered in the body of this article), or the scattering
of gravitational wave scattering into black holes
(for the ingoing null coordinates considered in
Appendix \ref{sec:ingoing_formalism}).

We have only considered a symmetric hyperbolic formulation of one kind
of single-null coordinate system, namely affine-null coordinates.
More general single null coordinates can be written as
\begin{align}
   ds^2
   =&
   Ve^{2B}du^2
   +
   2e^{2B}dudr
   \nonumber\\&
   -
   h_{ab}\left(d\theta^a+W^adu\right)\left(d\theta^b+W^bdu\right)
   ,
\end{align}
where we have introduced the function $B(u,r,\theta^a)$
(c.f. \cite{Giannakopoulos:2020dih}).
This form of the metric
captures a wide variety of single-null coordinate systems, e.g.
when $B=0$ we recover affine-null coordinates, and when 
$\mathrm{det}h_{ab} = r^2 \mathrm{det}s_{ab}$, where $s_{ab}$ is
the metric of the unit two-sphere, we recover the classical Bondi-Sachs
coordinates.
We leave to future work the construction of 
strongly hyperbolic formulations of the Einstein equations for this
more general set of single-null coordinates.
\begin{acknowledgments}
   We thank David Hilditch for helpful conversations
   about hyperbolic reductions and the Einstein equations, and for
   reviewing an earlier version of this text.
   We thank Elena Giorgi, Nicholas Loutrel, and Frans Pretorius
   for conversations about the Newman-Penrose formalism, and for
   discussions on an earlier version of this text.
\end{acknowledgments}

\section*{Data availability}
Data sharing is not applicable to this article as no new data were created
or analyzed in this study.
\appendix
\section{\label{sec:ingoing_formalism}
   A symmetric hyperbolic formulation of the Einstein equations
   in ingoing affine-null coordinates
   }
For completeness, here we present a symmetric hyperbolic formulation
of the Einstein equations in ingoing affine-null coordinates:
\begin{align}
\label{eq:ingoing_affine_null}
   ds^2
   =&
      Vdv^2
   -  2dvdr
   \nonumber\\&
   -  h_{ab}\left(d\theta^a+W^adv\right)\left(d\theta^b+W^bdv\right)
   .
\end{align}
Here $v$ is the null coordinate for ingoing null hypersurfaces.
While outgoing affine-null coordinates are specially adapted
to outgoing gravitational radiation, ingoing affine-null coordinates
find most of their use in studies of black hole/black brane horizons.
For example, ingoing affine-null coordinates have been widely used
in numerical simulations of asymptotically AdS
spacetimes that contain black holes/black branes \cite{Chesler:2013lia}.

We choose the following tetrad 
\begin{subequations}
\label{eq:ingoing_null_tetrad}
\begin{align}
\label{eq:ingoing_n_tetrad}
   n^{\mu}\partial_{\mu}
   =&
   -  \partial_r
   ,\\
\label{eq:ingoing_l_tetrad}
   l^{\mu}\partial_{\mu}
   =&
      \partial_v
   +  P\partial_r
   +  R^a\partial_a
   ,\\
\label{eq:ingoing_m_tetrad}
   m^{\mu}\partial_{\mu}
   =&
      Q\partial_r
   +  S^a\partial_a
   .
\end{align}
\end{subequations}
where $Q$ and $S^a$ are complex.
We use the six tetrad degrees of freedom to set
\begin{align}
   \nu=\tau=\gamma=0
   . 
\end{align}
In other words, we use our six tetrad degrees of freedom to
make $n^{\mu}$ geodesic, and to parallel propagate
$l^{\mu}$ and $m^{\mu}$ along that geodesic:
$
   n^{\mu}\nabla_{\mu}n^{\nu}
   =
   n^{\mu}\nabla_{\mu}l^{\nu}
   =
   n^{\mu}\nabla_{\mu}m^{\nu}
   =0
   .
$
We can find the coefficients for $l^{\mu}$ and $m^{\mu}$
by using the commutation relations\eqref{eq:commutation_relations}:
\begin{subequations}
\label{eq:ingoing_tetrad_evo}
\begin{align}
   \Delta l^{\mu}
   -
   \left(\epsilon+\bar{\epsilon}\right)n^{\mu}
   +
   \pi m^{\mu}
   +
   \bar{\pi}\bar{m}^{\mu}
   =&
   0
   ,\\
   \Delta m^{\mu}
   -
   \left(\bar{\alpha}+\beta\right)n^{\mu}
   +
   \mu m^{\mu}
   +
   \bar{\lambda}\bar{m}^{\mu}
   =&
   0
   .
\end{align}
\end{subequations}
Writing Eq.~\eqref{eq:ingoing_tetrad_evo} in terms of tetrad
components, we have
\begin{subequations}
\begin{align}
   DP
   +  \left(\epsilon+\bar{\epsilon}\right)
   +  \pi Q
   +  \bar{\pi}\bar{Q}
   =&
   0,\\
   DR^a
   +  \pi S^a
   +  \bar{\pi}\bar{S}^a
   =&
   0,\\
   DQ
   +  \left(\bar{\alpha}+\beta\right)
   +  \mu Q
   +  \bar{\lambda}\bar{Q}
   =&
   0,\\
   DS^a
   +  \mu S^a
   +  \bar{\lambda}\bar{S}^a
   =&
   0
   .
\end{align}
\end{subequations}

From the Ricci rotation identities~\eqref{eq:riem-all},
we obtain the following evolution equations for the nonzero
Ricci rotation coefficients
\begin{subequations}
\label{eq:ingoing_ricci}
\begin{align}
   \Delta\lambda
   +  \left(\mu+\bar{\mu}\right)\lambda
   +  \Psi_4
   =&
   0
   ,\\
   \Delta\mu
   +  \mu^2
   +  \lambda\bar{\lambda}
   =&
   0
   ,\\
   \Delta\pi
   +  \mu\pi
   +  \lambda\bar{\pi}
   +  \Psi_3
   =&
   0
   ,\\
   \Delta\rho
   +  \bar{\mu}\rho
   +  \sigma\lambda
   +  \Psi_2
   =&
   0
   ,\\
   \Delta\sigma
   +  \mu\sigma
   +  \bar{\lambda}\rho
   =&
   0
   ,\\
   \Delta\kappa
   +  \bar{\pi}\rho
   +  \pi\sigma
   +  \Psi_1
   =&
   0
   ,
   \\
   \Delta\alpha
   +  \bar{\mu}\alpha
   +  \lambda\beta
   +  \Psi_3
   =&
   0
   ,\\
   \Delta\beta
   +  \mu\beta
   +  \bar{\lambda}\alpha
   =&
   0
   ,\\
   \Delta\epsilon
   +  \bar{\pi}\alpha
   +  \pi\beta
   +  \Psi_2
   =&
   0
   .
\end{align}
\end{subequations}
We rewrite the Bianchi identities to obtain the following evolution
system for the Weyl scalars:
\begin{subequations}
\begin{align}
   \left(\Delta+\mu\right)\Psi_0
   -
   \edth\Psi_1
   -
   3\sigma\Psi_2
   &=
   0
   ,\\
   \left(\Delta+\thorn+2\mu-4\rho\right)\Psi_1
   -
   \left(\edth-3\kappa\right)\Psi_2
   \nonumber\\
   -
   \left(\edth'+\pi\right)\Psi_0
   -2\sigma\Psi_3
   &=
   0
   ,\\
   \left(\Delta+\thorn+3\mu-3\rho\right)\Psi_2
   -
   \left(\edth-2\kappa\right)\Psi_3
   &\nonumber\\
   -
   \left(\edth'+2\pi\right)\Psi_1
   +
   \lambda\Psi_0
   -
   \sigma\Psi_4
   &=
   0
   ,\\
   \left(\Delta+\thorn+4\mu-2\rho\right)\Psi_3
   -
   \left(\edth-\kappa\right)\Psi_4
   &\nonumber\\
   -
   \left(\edth'+3\pi\right)\Psi_2
   +
   2\lambda\Psi_1
   &=
   0
   ,\\
   \left(\thorn-\rho\right)\Psi_4
   -
   \left(\edth'+4\pi\right)\Psi_3
   +
   3\lambda\Psi_2
   &=
   0
   .
\end{align}
\end{subequations}
As in Sec.~\ref{sec:outgoing_eom},
we obtained this system by adding together
(\eqref{eq:bianchi-1}+\eqref{eq:bianchi-6}),
(\eqref{eq:bianchi-2}+\eqref{eq:bianchi-7}),
and
(\eqref{eq:bianchi-3}+\eqref{eq:bianchi-8}).

Similarly to what we did in Sec.~\ref{sec:Initial data}
we can specify initial data by setting all the Newman-Penrose
scalar to be equal to some exact, background solution
at $r=r_{max}$ on the $v=v_i$ initial hypersurface.
We can then construct outgoing radiation on that slice by freely
specifying the real and imaginary parts of ${}^+\Psi_4$,
and then integrating \emph{inwards} in $r$ from $r=r_{max}$
the following constraint equations for the Weyl scalars
(which are taken from the Bianchi identities 
\ref{eq:bianchi-5}-\ref{eq:bianchi-8})
\begin{subequations}
\begin{align}
   \left(\Delta+\mu\right)\Psi_0
   -
   \edth\Psi_1
   -
   3\sigma\Psi_2
   =&
   0
   ,\\
   \left(\Delta+2\mu\right)\Psi_1
   -
   \edth\Psi_2
   -
   2\sigma\Psi_3
   =&
   0
   ,\\
   \left(\Delta+3\mu\right)\Psi_2
   -
   \edth\Psi_3
   -
   \sigma\Psi_4
   =&
   0
   ,\\
   \left(\Delta+4\mu\right)\Psi_3
   -
   \edth\Psi_4
   =&
   0
   .
\end{align}
\end{subequations}
the tetrad components, and the Ricci rotation coefficients
from their equations of motion Eqs.
\eqref{eq:ingoing_tetrad_evo}
and \eqref{eq:ingoing_ricci}, respectively.
To evolve in time, we could set all the Newman-Penrose scalars
to be equal to their background value at $r=r_{max}$, and
specify $\Psi_4={}^{(bkgrd)}\Psi_4+{}^+\Psi_4$, where
${}^+\Psi_4$ is specified over a compact region of the initial
data surface. One potential advantage of the
ingoing formulation is that if we compactify in $r$ to spatial infinity,
then our boundary data at spatial infinity will always be consistent.
With regards to what happens at the origin ($r=0$), typically
ingoing affine-null coordinates are used to study black hole/black brane
spacetimes, and in those cases
the domain only needs to extend to the black hole horizon
(for more discussion see e.g. \cite{Chesler:2013lia}).
\section{\label{sec:hyperbolicity_review}
   A brief review of weak, strong, and symmetric hyperbolic systems 
   }
   For completeness, we review the notions of weak, strong, and symmetric
   hyperbolic systems; for a more in-depth account see, e.g.
   \cite{kreiss1989initial,Sarbach:2012pr,Hilditch:2013sba}.
   We begin with a system of evolution PDE of the form
   \begin{align}
      \label{eq:general_system}
      \hat{A}\partial_t{\bf u} 
      + 
      \hat{B}^i\left(t,{\bf x},{\bf u}\right)\partial_i{\bf u}
      +
      {\bf F}\left(t,{\bf x},{\bf u}\right)
      =
      0
      ,
   \end{align}
   where ${\bf u}$ is the state vector,
   $\partial_i\equiv\partial/\partial x^i$, and we sum over the indices $i$.
   The system \eqref{eq:general_system} is \emph{weakly hyperbolic}
   if the \emph{principal symbol} 
   \begin{align}
      \mathcal{P}\left(\xi\right)\equiv\hat{A}^{-1}\hat{B}^i\xi_i
   \end{align}
   has all real eigenvalues for all unit spatial vectors $\xi_i$.
   It is \emph{strongly hyperbolic} if for every unit spatial vector
   $\xi_i$ the principal symbol has a complete set of eigenvectors,
   and that there exists a constant $K$ independent of $\xi_i$
   such that 
   \begin{align}
      \left|\hat{T}_{\xi}\right| 
      + 
      \left|\hat{T}^{-1}_{\xi}\right|
      \leq 
      K
      ,
   \end{align}
   where the columns of the matrix $\hat{T}_{\xi}$
   are given by the eigenvectors of $\mathcal{P}\left(\xi\right)$.
   The system \eqref{eq:general_system} is
   \emph{symmetric hyperbolic} if there is a Hermitian positive
   definite symmetrizer $H$ such that $HA^i\xi_i$ is Hermitian
   for every unit spatial vector $\xi_i$.
   Symmetric hyperbolicity implies strong hyperbolicity,
   and strong hyperbolicity implies weak hyperbolicity,
   but the converse of these statements do not hold.

   In this article we consider systems of PDE that take the slightly
   more general form
   \begin{align}
      \label{eq:covariant_eqn}
      \hat{A}^{\mu}\left(t,{\bf x},{\bf u}\right)\partial_{\mu}{\bf u}
      +
      {\bf F}\left(t,{\bf x},{\bf u}\right)
      =
      0
      ,
   \end{align}
   where $x^\mu$ ranges over space and time.
   If there is a timelike vector $t^{\mu}$ such that then the matrix
   \begin{align}
      \hat{A}^{\mu}t_{\mu}
      ,
   \end{align}
   is positive definite, we can then perform a coordinate transform to
   make \eqref{eq:covariant_eqn} take the more classical
   form \eqref{eq:general_system} (in particular, transform coordinates
   so that $t_{\mu}=\delta^0_{\mu}$; then 
   $\hat{A}\equiv \hat{A}^0$, and $\hat{B}^i\equiv \hat{A}^i$
   in the new coordinate system).
   We conclude that in order to show that systems of PDE that take
   the form \eqref{eq:covariant_eqn} are
   symmetric hyperbolic, it is sufficient to show that
   $\hat{A}^{\mu}\xi_{\mu}$ is Hermitian for all $\xi_{\mu}$, 
   and that there is a timelike vector $t^{\mu}$ such that
   $\hat{A}^{\mu}t_{\mu}$ is positive definite.
\begin{widetext}
\section{\label{sec:review_NP_formalism}
   The Newman-Penrose and Geroch-Held-Penrose formalisms
   }
The Newman-Penrose \cite{Newman_Penrose_paper}
and Geroch-Held-Penrose \cite{GHP_paper} are tetrad based reformulations
of the Einstein equations.
For reference, we list the main
Newman-Penrose equations that we use in this article.
We note that we only consider the \emph{vacuum} Einstein equations,
which in terms of the metric form the following system of equations:
\begin{align}
   R_{\mu\nu}
   =
   0
   ,
\end{align}
where $R_{\mu\nu}$ is the Ricci tensor.
For a more complete description of the Newman-Penrose and
Geroch-Held-Penrose formalisms,
see \cite{Newman_Penrose_paper,GHP_paper,Chandrasekhar_bh_book};
our notation follows Newman and Penrose\cite{Newman_Penrose_paper}
(see also Chandrasekhar\cite{Chandrasekhar_bh_book}).
As is standard, we denote the null tetrad as $n^{\mu}, l^{\mu}, m^{\mu}$,
where $m^{\mu}$ is a complex null vector, $n^{\mu}$ is an ingoing null vector,
and $l^{\mu}$ is an outgoing null vector. 
The metric then is
\begin{align}
   g_{\mu\nu}
   =
   n_{\mu}l_{\nu}
   +
   l_{\mu}n_{\nu}
   -
   m_{\mu}\bar{m}_{\nu}
   -
   \bar{m}_{\mu}m_{\nu}
   .
\end{align}
The derivative operators are 
\begin{align}
   D\equiv l^{\mu}\partial_{\mu},
   \qquad
   \Delta\equiv n^{\mu}\partial_{\mu},
   \qquad
   \delta\equiv m^{\mu}\partial_{\mu}
   .
\end{align}
   The derivative commutation relations are
\begin{subequations}
\label{eq:commutation_relations}
\begin{align}
   \Delta D - D \Delta
   =&
   \left(\gamma+\bar{\gamma}\right)D
   +
   \left(\epsilon+\bar{\epsilon}\right)\Delta
   -
   \left(\bar{\tau}+\pi\right)\delta
   -
   \left(\tau+\bar{\pi}\right)\bar{\delta}
   ,\\
   \Delta D - D \delta
   =&
   \left(\bar{\alpha}+\beta-\bar{\pi}\right)D
   +
   \kappa\Delta
   -
   \left(\bar{\rho}+\epsilon-\bar{\epsilon}\right)\delta
   -
   \sigma\bar{\delta}
   ,\\
   \delta\Delta - \Delta\delta
   =&
   -\bar{\nu}D
   +
   \left(\tau-\bar{\alpha}-\beta\right)\Delta
   +
   \left(\mu-\gamma+\bar{\gamma}\right)\delta
   +
   \bar{\lambda}\bar{\delta}
   ,\\
   \bar{\delta}\delta-\delta\bar{\delta}
   =&
   \left(\bar{\mu}-\mu\right)D
   +
   \left(\bar{\rho}-\rho\right)\Delta
   +
   \left(\alpha-\bar{\beta}\right)\delta
   +
   \left(\beta-\bar{\alpha}\right)\bar{\delta}
   .
\end{align}
\end{subequations}
The vacuum Ricci rotation identities are
\begin{subequations}
\label{eq:riem-all}
\begin{align}
\label{eq:riem-1}
{D}\rho - \bar{{\delta}} \kappa &= (\rho^{2} + \sigma \bar{\sigma}) + \rho (\epsilon + \bar{\epsilon}) - \bar{\kappa} \tau - \kappa (3\alpha + \bar{\beta} - \pi),
\\
\label{eq:riem-2}
{D}\sigma - {\delta}\kappa &= \sigma(\rho + \bar{\rho} + 3 \epsilon - \bar{\epsilon}) - \kappa(\tau - \bar{\pi} + \bar{\alpha} + 3\beta) + \Psi_{0}
\\
\label{eq:riem-3}
{D}\tau - {\Delta}\kappa &= \rho(\tau + \bar{\pi}) + \sigma(\bar{\tau} + \pi) + \tau(\epsilon - \bar{\epsilon}) - \kappa(3\gamma + \bar{\gamma}) + \Psi_{1} 
\\
\label{eq:riem-4}
{D}\alpha - \bar{{\delta}}\epsilon &= \alpha (\rho + \bar{\epsilon} - 2 \epsilon) + \beta \bar{\sigma} - \bar{\beta} \epsilon - \kappa \lambda - \bar{\kappa} \gamma + \pi (\epsilon + \rho)
\\
\label{eq:riem-5}
{D}\beta - {\delta}\epsilon &= \sigma(\alpha + \pi) + \beta(\bar{\rho}-\bar{\epsilon}) - \kappa(\mu + \gamma) - \epsilon(\bar{\alpha}-\bar{\pi}) + \Psi_{1}
\\
\label{eq:riem-6}
{D}\gamma - {\Delta}\epsilon &= \alpha(\tau+\bar{\pi}) + \beta(\bar{\tau}+\pi) - \gamma(\epsilon + \bar{\epsilon}) - \epsilon(\gamma + \bar{\gamma}) + \tau \pi - \nu \kappa + \Psi_{2}
\\
\label{eq:riem-7}
{D}\lambda - \bar{{\delta}}\pi &= (\rho \lambda + \bar{\sigma} \mu) + \pi (\pi + \alpha - \bar{\beta}) - \nu \bar{\kappa} - \lambda (3\epsilon - \bar{\epsilon}) 
\\
\label{eq:riem-8}
{D}\mu - {\delta}\pi &= (\bar{\rho}\mu + \sigma \lambda) + \pi(\bar{\pi} - \bar{\alpha} + \beta) - \mu(\epsilon+\bar{\epsilon}) - \nu \kappa + \Psi_{2}
\\
\label{eq:riem-9}
{D}\nu - {\Delta}\pi &= \mu(\pi + \bar{\tau}) + \lambda(\bar{\pi} + \tau) + \pi(\gamma-\bar{\gamma}) - \nu(3\epsilon+\bar{\epsilon}) + \Psi_{3}
\\
\label{eq:riem-10}
{\Delta}\lambda - \bar{{\delta}}\nu &= -\lambda(\mu + \bar{\mu} + 3\gamma - \bar{\gamma}) + \nu(3\alpha + \bar{\beta} + \pi - \bar{\tau}) - \Psi_{4}
\\
\label{eq:riem-11}
{\delta}\rho - \bar{{\delta}}\sigma &= \rho(\bar{\alpha} + \beta) - \sigma(3\alpha-\bar{\beta}) + \tau(\rho-\bar{\rho}) + \kappa(\mu-\bar{\mu}) - \Psi_{1}
\\
\label{eq:riem-12}
{\delta}\alpha - \bar{{\delta}}\beta &= \mu \rho - \lambda \sigma + \alpha \bar{\alpha} + \beta \bar{\beta} - 2\alpha \beta + \gamma(\rho - \bar{\rho}) + \epsilon(\mu - \bar{\mu}) - \Psi_{2}
\\
\label{eq:riem-13}
{\delta}\lambda - \bar{{\delta}}\mu &= \nu(\rho-\bar{\rho}) + \pi (\mu - \bar{\mu}) + \mu (\alpha + \bar{\beta}) + \lambda(\bar{\alpha} - 3\beta) - \Psi_{3}
\\
\label{eq:riem-14}
{\delta}\nu - {\Delta}\mu &= (\mu^{2} + \lambda \bar{\lambda}) + \mu (\gamma + \bar{\gamma}) - \bar{\nu} \pi + \nu (\tau - 3 \beta - \bar{\alpha})
\\
\label{eq:riem-15}
{\delta}\gamma - {\Delta}\beta &= \gamma(\tau - \bar{\alpha} - \beta) + \mu \tau - \sigma \nu - \epsilon \bar{\nu} - \beta (\gamma - \bar{\gamma} - \mu) + \alpha \bar{\lambda}
\\
\label{eq:riem-16}
{\delta}\tau - {\Delta}\sigma &= (\mu \sigma + \bar{\lambda}\rho) + \tau (\tau + \beta - \bar{\alpha}) - \sigma (3\gamma - \bar{\gamma}) - \kappa \bar{\nu}
\\
\label{eq:riem-17}
{\Delta}\rho - \bar{{\delta}}\tau &= -\rho \bar{\mu} + \sigma \lambda + \tau(\bar{\beta} - \alpha - \bar{\tau}) + \rho(\gamma + \bar{\gamma}) + \nu \kappa - \Psi_{2}
\\
\label{eq:riem-18}
{\Delta}\alpha - \bar{{\delta}}\gamma &= \nu(\rho + \epsilon) - \lambda(\tau+\beta) + \alpha(\bar{\gamma} - \bar{\mu}) + \gamma(\bar{\beta} - \bar{\tau}) - \Psi_{3}
\end{align}
\end{subequations}
The vacuum Bianchi identities are
\begin{subequations}
\label{eq:bianchi-all}
\begin{align}
\label{eq:bianchi-1}
   -  \bar{{\delta}} \Psi_{0} + {D}\Psi_{1} 
   +  (4\alpha - \pi) \Psi_{0} - 2(2\rho + \epsilon) \Psi_{1} 
   + 3 \kappa \Psi_{2} + 
   &= 0,
\\
\label{eq:bianchi-2}
\bar{{\delta}}\Psi_{1} - {D} \Psi_{2} - \lambda \Psi_{0} + 2(\pi - \alpha) \Psi_{1} + 3 \rho \Psi_{2} - 2 \kappa \Psi_{3} &= 0,
\\
\label{eq:bianchi-3}
-\bar{{\delta}}\Psi_{2} + {D} \Psi_{3} + 2 \lambda \Psi_{1} - 3 \pi \Psi_{2} + 2 (\epsilon - \rho) \Psi_{3} + \kappa \Psi_{4} + &= 0,
\\
\label{eq:bianchi-4}
\bar{{\delta}}\Psi_{3} - {D} \Psi_{4} - 3 \lambda \Psi_{2} + 2 (2\pi + \alpha) \Psi_{3} - (4\epsilon - \rho) \Psi_{4} &= 0,
\\
\label{eq:bianchi-5}
-{\Delta} \Psi_{0} + {\delta}\Psi_{1} + (4\gamma - \mu)\Psi_{0} - 2(2\tau + \beta) \Psi_{1} + 3\sigma \Psi_{2} &= 0,
\\
\label{eq:bianchi-6}
-{\Delta} \Psi_{1} + {\delta} \Psi_{2} + \nu \Psi_{0} + 2(\gamma-\mu)\Psi_{1} - 3\tau \Psi_{2} + 2\sigma \Psi_{3} &= 0,
\\
\label{eq:bianchi-7}
-{\Delta} \Psi_{2} + {\delta} \Psi_{3} + 2 \nu \Psi_{1} - 3\mu \Psi_{2} + 2 (\beta - \tau) \Psi_{3} + \sigma \Psi_{4} &= 0,
\\
\label{eq:bianchi-8}
-{\Delta} \Psi_{3} + {\delta} \Psi_{4} + 3 \nu \Psi_{2} - 2(\gamma + 2\mu) \Psi_{3} - (\tau - 4\beta)\Psi_{4} &= 0,
\end{align}
\end{subequations}
Geroch, Held, and Penrose \cite{GHP_paper} showed that 
the derivative operators $D,\Delta,\delta$, when acting on the Newman-Penrose
scalars are generically not invariant under tetrad rotations.
They classified scalars $f$ as having weights $(p,q)$ if under
the rescaling (here $c$ is a complex field)
$l^{\mu}\to c\bar{c} l^{\mu}$, 
$n^{\mu}\to c^{-1}\bar{c}^{-1} n^{\mu}$,
$m^{\mu}\to c\bar{c}^{-1}m^{\mu}$, it transforms as $f\to c^p\bar{c}^q f$.
For a scalar field $f$ of weight $(p,q)$,
they defined the following tetrad rotation covariant operators
\begin{align}
   \thorn f 
   \equiv
   \left(D-p\epsilon-q\bar{\epsilon}\right)f
   ,\qquad
   \thorn'f 
   \equiv 
   \left(\Delta-p\gamma-q\bar{\gamma}\right)f
   ,\qquad
   \edth f 
   \equiv 
   \left(\delta-p\beta-q\bar{\alpha}\right)f
   ,\qquad
   \edth'f 
   \equiv 
   \left(\bar{\delta}-p\alpha-q\bar{\beta}\right)f
   .
\end{align}
\end{widetext}
\bibliography{references.bib}

\begin{thebibliography}{26}%
\makeatletter
\providecommand \@ifxundefined [1]{%
 \@ifx{#1\undefined}
}%
\providecommand \@ifnum [1]{%
 \ifnum #1\expandafter \@firstoftwo
 \else \expandafter \@secondoftwo
 \fi
}%
\providecommand \@ifx [1]{%
 \ifx #1\expandafter \@firstoftwo
 \else \expandafter \@secondoftwo
 \fi
}%
\providecommand \natexlab [1]{#1}%
\providecommand \enquote  [1]{``#1''}%
\providecommand \bibnamefont  [1]{#1}%
\providecommand \bibfnamefont [1]{#1}%
\providecommand \citenamefont [1]{#1}%
\providecommand \href@noop [0]{\@secondoftwo}%
\providecommand \href [0]{\begingroup \@sanitize@url \@href}%
\providecommand \@href[1]{\@@startlink{#1}\@@href}%
\providecommand \@@href[1]{\endgroup#1\@@endlink}%
\providecommand \@sanitize@url [0]{\catcode `\\12\catcode `\$12\catcode
  `\&12\catcode `\#12\catcode `\^12\catcode `\_12\catcode `\%12\relax}%
\providecommand \@@startlink[1]{}%
\providecommand \@@endlink[0]{}%
\providecommand \url  [0]{\begingroup\@sanitize@url \@url }%
\providecommand \@url [1]{\endgroup\@href {#1}{\urlprefix }}%
\providecommand \urlprefix  [0]{URL }%
\providecommand \Eprint [0]{\href }%
\providecommand \doibase [0]{http://dx.doi.org/}%
\providecommand \selectlanguage [0]{\@gobble}%
\providecommand \bibinfo  [0]{\@secondoftwo}%
\providecommand \bibfield  [0]{\@secondoftwo}%
\providecommand \translation [1]{[#1]}%
\providecommand \BibitemOpen [0]{}%
\providecommand \bibitemStop [0]{}%
\providecommand \bibitemNoStop [0]{.\EOS\space}%
\providecommand \EOS [0]{\spacefactor3000\relax}%
\providecommand \BibitemShut  [1]{\csname bibitem#1\endcsname}%
\let\auto@bib@innerbib\@empty
\bibitem [{\citenamefont {Bondi}(1960)}]{Bondi:1960jsa}%
  \BibitemOpen
  \bibfield  {author} {\bibinfo {author} {\bibfnamefont {H.}~\bibnamefont
  {Bondi}},\ }\href {\doibase 10.1038/186535a0} {\bibfield  {journal} {\bibinfo
   {journal} {Nature}\ }\textbf {\bibinfo {volume} {186}},\ \bibinfo {pages}
  {535} (\bibinfo {year} {1960})}\BibitemShut {NoStop}%
\bibitem [{\citenamefont {Sachs}\ and\ \citenamefont
  {Bondi}(1962)}]{doi:10.1098/rspa.1962.0206}%
  \BibitemOpen
  \bibfield  {author} {\bibinfo {author} {\bibfnamefont {R.~K.~.}\ \bibnamefont
  {Sachs}}\ and\ \bibinfo {author} {\bibfnamefont {H.}~\bibnamefont {Bondi}},\
  }\href {\doibase 10.1098/rspa.1962.0206} {\bibfield  {journal} {\bibinfo
  {journal} {Proceedings of the Royal Society of London. Series A. Mathematical
  and Physical Sciences}\ }\textbf {\bibinfo {volume} {270}},\ \bibinfo {pages}
  {103} (\bibinfo {year} {1962})},\ \Eprint
  {http://arxiv.org/abs/https://royalsocietypublishing.org/doi/pdf/10.1098/rspa.1962.0206}
  {https://royalsocietypublishing.org/doi/pdf/10.1098/rspa.1962.0206}
  \BibitemShut {NoStop}%
\bibitem [{\citenamefont {Bondi}\ \emph {et~al.}(1962)\citenamefont {Bondi},
  \citenamefont {Van~der Burg},\ and\ \citenamefont
  {Metzner}}]{doi:10.1098/rspa.1962.0161}%
  \BibitemOpen
  \bibfield  {author} {\bibinfo {author} {\bibfnamefont {H.}~\bibnamefont
  {Bondi}}, \bibinfo {author} {\bibfnamefont {M.~G.~J.}\ \bibnamefont {Van~der
  Burg}}, \ and\ \bibinfo {author} {\bibfnamefont {A.~W.~K.}\ \bibnamefont
  {Metzner}},\ }\href {\doibase 10.1098/rspa.1962.0161} {\bibfield  {journal}
  {\bibinfo  {journal} {Proceedings of the Royal Society of London. Series A.
  Mathematical and Physical Sciences}\ }\textbf {\bibinfo {volume} {269}},\
  \bibinfo {pages} {21} (\bibinfo {year} {1962})},\ \Eprint
  {http://arxiv.org/abs/https://royalsocietypublishing.org/doi/pdf/10.1098/rspa.1962.0161}
  {https://royalsocietypublishing.org/doi/pdf/10.1098/rspa.1962.0161}
  \BibitemShut {NoStop}%
\bibitem [{\citenamefont {Tamburino}\ and\ \citenamefont
  {Winicour}(1966)}]{Tamburino:1966zz}%
  \BibitemOpen
  \bibfield  {author} {\bibinfo {author} {\bibfnamefont {L.~A.}\ \bibnamefont
  {Tamburino}}\ and\ \bibinfo {author} {\bibfnamefont {J.~H.}\ \bibnamefont
  {Winicour}},\ }\href {\doibase 10.1103/PhysRev.150.1039} {\bibfield
  {journal} {\bibinfo  {journal} {Phys. Rev.}\ }\textbf {\bibinfo {volume}
  {150}},\ \bibinfo {pages} {1039} (\bibinfo {year} {1966})}\BibitemShut
  {NoStop}%
\bibitem [{\citenamefont {Bishop}\ \emph {et~al.}(1996)\citenamefont {Bishop},
  \citenamefont {Gomez}, \citenamefont {Lehner},\ and\ \citenamefont
  {Winicour}}]{Bishop:1996gt}%
  \BibitemOpen
  \bibfield  {author} {\bibinfo {author} {\bibfnamefont {N.~T.}\ \bibnamefont
  {Bishop}}, \bibinfo {author} {\bibfnamefont {R.}~\bibnamefont {Gomez}},
  \bibinfo {author} {\bibfnamefont {L.}~\bibnamefont {Lehner}}, \ and\ \bibinfo
  {author} {\bibfnamefont {J.}~\bibnamefont {Winicour}},\ }\href {\doibase
  10.1103/PhysRevD.54.6153} {\bibfield  {journal} {\bibinfo  {journal} {Phys.
  Rev. D}\ }\textbf {\bibinfo {volume} {54}},\ \bibinfo {pages} {6153}
  (\bibinfo {year} {1996})},\ \Eprint {http://arxiv.org/abs/gr-qc/9705033}
  {arXiv:gr-qc/9705033} \BibitemShut {NoStop}%
\bibitem [{\citenamefont {Handmer}\ \emph {et~al.}(2016)\citenamefont
  {Handmer}, \citenamefont {Szil\'agyi},\ and\ \citenamefont
  {Winicour}}]{Handmer:2016mls}%
  \BibitemOpen
  \bibfield  {author} {\bibinfo {author} {\bibfnamefont {C.~J.}\ \bibnamefont
  {Handmer}}, \bibinfo {author} {\bibfnamefont {B.}~\bibnamefont {Szil\'agyi}},
  \ and\ \bibinfo {author} {\bibfnamefont {J.}~\bibnamefont {Winicour}},\
  }\href {\doibase 10.1088/0264-9381/33/22/225007} {\bibfield  {journal}
  {\bibinfo  {journal} {Class. Quant. Grav.}\ }\textbf {\bibinfo {volume}
  {33}},\ \bibinfo {pages} {225007} (\bibinfo {year} {2016})},\ \Eprint
  {http://arxiv.org/abs/1605.04332} {arXiv:1605.04332 [gr-qc]} \BibitemShut
  {NoStop}%
\bibitem [{\citenamefont {Winicour}(2009)}]{Winicour:2008vpn}%
  \BibitemOpen
  \bibfield  {author} {\bibinfo {author} {\bibfnamefont {J.}~\bibnamefont
  {Winicour}},\ }\href {\doibase 10.12942/lrr-2009-3} {\bibfield  {journal}
  {\bibinfo  {journal} {Living Rev. Rel.}\ }\textbf {\bibinfo {volume} {12}},\
  \bibinfo {pages} {3} (\bibinfo {year} {2009})},\ \Eprint
  {http://arxiv.org/abs/0810.1903} {arXiv:0810.1903 [gr-qc]} \BibitemShut
  {NoStop}%
\bibitem [{\citenamefont {Chesler}\ and\ \citenamefont
  {Yaffe}(2014)}]{Chesler:2013lia}%
  \BibitemOpen
  \bibfield  {author} {\bibinfo {author} {\bibfnamefont {P.~M.}\ \bibnamefont
  {Chesler}}\ and\ \bibinfo {author} {\bibfnamefont {L.~G.}\ \bibnamefont
  {Yaffe}},\ }\href {\doibase 10.1007/JHEP07(2014)086} {\bibfield  {journal}
  {\bibinfo  {journal} {JHEP}\ }\textbf {\bibinfo {volume} {07}},\ \bibinfo
  {pages} {086} (\bibinfo {year} {2014})},\ \Eprint
  {http://arxiv.org/abs/1309.1439} {arXiv:1309.1439 [hep-th]} \BibitemShut
  {NoStop}%
\bibitem [{\citenamefont {Giannakopoulos}\ \emph {et~al.}(2020)\citenamefont
  {Giannakopoulos}, \citenamefont {Hilditch},\ and\ \citenamefont
  {Zilhao}}]{Giannakopoulos:2020dih}%
  \BibitemOpen
  \bibfield  {author} {\bibinfo {author} {\bibfnamefont {T.}~\bibnamefont
  {Giannakopoulos}}, \bibinfo {author} {\bibfnamefont {D.}~\bibnamefont
  {Hilditch}}, \ and\ \bibinfo {author} {\bibfnamefont {M.}~\bibnamefont
  {Zilhao}},\ }\href {\doibase 10.1103/PhysRevD.102.064035} {\bibfield
  {journal} {\bibinfo  {journal} {Phys. Rev. D}\ }\textbf {\bibinfo {volume}
  {102}},\ \bibinfo {pages} {064035} (\bibinfo {year} {2020})},\ \Eprint
  {http://arxiv.org/abs/2007.06419} {arXiv:2007.06419 [gr-qc]} \BibitemShut
  {NoStop}%
\bibitem [{\citenamefont {Friedrich}(1981)}]{doi:10.1098/rspa.1981.0045}%
  \BibitemOpen
  \bibfield  {author} {\bibinfo {author} {\bibfnamefont {H.}~\bibnamefont
  {Friedrich}},\ }\href {\doibase 10.1098/rspa.1981.0045} {\bibfield  {journal}
  {\bibinfo  {journal} {Proceedings of the Royal Society of London. A.
  Mathematical and Physical Sciences}\ }\textbf {\bibinfo {volume} {375}},\
  \bibinfo {pages} {169} (\bibinfo {year} {1981})},\ \Eprint
  {http://arxiv.org/abs/https://royalsocietypublishing.org/doi/pdf/10.1098/rspa.1981.0045}
  {https://royalsocietypublishing.org/doi/pdf/10.1098/rspa.1981.0045}
  \BibitemShut {NoStop}%
\bibitem [{\citenamefont {Luk}(2011)}]{Luk:2011vf}%
  \BibitemOpen
  \bibfield  {author} {\bibinfo {author} {\bibfnamefont {J.}~\bibnamefont
  {Luk}},\ }\href@noop {} {\  (\bibinfo {year} {2011})},\ \Eprint
  {http://arxiv.org/abs/1107.0898} {arXiv:1107.0898 [gr-qc]} \BibitemShut
  {NoStop}%
\bibitem [{\citenamefont {Hilditch}\ \emph {et~al.}(2020)\citenamefont
  {Hilditch}, \citenamefont {Valiente~Kroon},\ and\ \citenamefont
  {Zhao}}]{Hilditch:2019maz}%
  \BibitemOpen
  \bibfield  {author} {\bibinfo {author} {\bibfnamefont {D.}~\bibnamefont
  {Hilditch}}, \bibinfo {author} {\bibfnamefont {J.~A.}\ \bibnamefont
  {Valiente~Kroon}}, \ and\ \bibinfo {author} {\bibfnamefont {P.}~\bibnamefont
  {Zhao}},\ }\href {\doibase 10.1007/s10714-020-02747-2} {\bibfield  {journal}
  {\bibinfo  {journal} {Gen. Rel. Grav.}\ }\textbf {\bibinfo {volume} {52}},\
  \bibinfo {pages} {99} (\bibinfo {year} {2020})},\ \Eprint
  {http://arxiv.org/abs/1911.00047} {arXiv:1911.00047 [gr-qc]} \BibitemShut
  {NoStop}%
\bibitem [{\citenamefont {Newman}\ and\ \citenamefont
  {Penrose}(1962)}]{Newman_Penrose_paper}%
  \BibitemOpen
  \bibfield  {author} {\bibinfo {author} {\bibfnamefont {E.}~\bibnamefont
  {Newman}}\ and\ \bibinfo {author} {\bibfnamefont {R.}~\bibnamefont
  {Penrose}},\ }\href {\doibase 10.1063/1.1724257} {\bibfield  {journal}
  {\bibinfo  {journal} {Journal of Mathematical Physics}\ }\textbf {\bibinfo
  {volume} {3}},\ \bibinfo {pages} {566} (\bibinfo {year} {1962})}\BibitemShut
  {NoStop}%
\bibitem [{Note1()}]{Note1}%
  \BibitemOpen
  \bibinfo {note} {We note that the possibility of a strongly hyperbolic
  formulation of the Einstein equations for Bondi-like coordinate systems
  existing for a tetrad-based formalism was suggested by Giannakopoulos et. al.
  \cite {Giannakopoulos:2020dih}.}\BibitemShut {Stop}%
\bibitem [{\citenamefont {Kreiss}\ and\ \citenamefont
  {Lorenz}(1989)}]{kreiss1989initial}%
  \BibitemOpen
  \bibfield  {author} {\bibinfo {author} {\bibfnamefont {H.}~\bibnamefont
  {Kreiss}}\ and\ \bibinfo {author} {\bibfnamefont {J.}~\bibnamefont
  {Lorenz}},\ }\href {https://books.google.co.uk/books?id=aCEq9TaC1tQC} {\emph
  {\bibinfo {title} {Initial-Boundary Value Problems and the Navier-Stokes
  Equations}}},\ ISSN\ (\bibinfo  {publisher} {Elsevier Science},\ \bibinfo
  {year} {1989})\BibitemShut {NoStop}%
\bibitem [{Note2()}]{Note2}%
  \BibitemOpen
  \bibinfo {note} {As is stressed in \cite {Giannakopoulos:2020dih} though,
  this does not imply that a numerical solution of a weakly hyperbolic system
  will necessarily exhibit any pathological behavior at a given \protect \emph
  {fixed} resolution}\BibitemShut {NoStop}%
\bibitem [{\citenamefont {Sarbach}\ and\ \citenamefont
  {Tiglio}(2012)}]{Sarbach:2012pr}%
  \BibitemOpen
  \bibfield  {author} {\bibinfo {author} {\bibfnamefont {O.}~\bibnamefont
  {Sarbach}}\ and\ \bibinfo {author} {\bibfnamefont {M.}~\bibnamefont
  {Tiglio}},\ }\href {\doibase 10.12942/lrr-2012-9} {\bibfield  {journal}
  {\bibinfo  {journal} {Living Rev. Rel.}\ }\textbf {\bibinfo {volume} {15}},\
  \bibinfo {pages} {9} (\bibinfo {year} {2012})},\ \Eprint
  {http://arxiv.org/abs/1203.6443} {arXiv:1203.6443 [gr-qc]} \BibitemShut
  {NoStop}%
\bibitem [{\citenamefont {Hilditch}(2013)}]{Hilditch:2013sba}%
  \BibitemOpen
  \bibfield  {author} {\bibinfo {author} {\bibfnamefont {D.}~\bibnamefont
  {Hilditch}},\ }\href {\doibase 10.1142/S0217751X13400150} {\bibfield
  {journal} {\bibinfo  {journal} {Int. J. Mod. Phys. A}\ }\textbf {\bibinfo
  {volume} {28}},\ \bibinfo {pages} {1340015} (\bibinfo {year} {2013})},\
  \Eprint {http://arxiv.org/abs/1309.2012} {arXiv:1309.2012 [gr-qc]}
  \BibitemShut {NoStop}%
\bibitem [{\citenamefont {Rendall}(1990)}]{doi:10.1098/rspa.1990.0009}%
  \BibitemOpen
  \bibfield  {author} {\bibinfo {author} {\bibfnamefont {A.~D.}\ \bibnamefont
  {Rendall}},\ }\href {\doibase 10.1098/rspa.1990.0009} {\bibfield  {journal}
  {\bibinfo  {journal} {Proceedings of the Royal Society of London. A.
  Mathematical and Physical Sciences}\ }\textbf {\bibinfo {volume} {427}},\
  \bibinfo {pages} {221} (\bibinfo {year} {1990})},\ \Eprint
  {http://arxiv.org/abs/https://royalsocietypublishing.org/doi/pdf/10.1098/rspa.1990.0009}
  {https://royalsocietypublishing.org/doi/pdf/10.1098/rspa.1990.0009}
  \BibitemShut {NoStop}%
\bibitem [{Note3()}]{Note3}%
  \BibitemOpen
  \bibinfo {note} {For a review which discusses earlier attempts to formulate a
  well-posed initial value problem for single-null formulations of the Einstein
  equations, see \cite {Winicour:2008vpn}.}\BibitemShut {Stop}%
\bibitem [{\citenamefont {Chandrasekhar}(2002)}]{Chandrasekhar_bh_book}%
  \BibitemOpen
  \bibfield  {author} {\bibinfo {author} {\bibfnamefont {S.}~\bibnamefont
  {Chandrasekhar}},\ }\href {https://cds.cern.ch/record/579245} {\emph
  {\bibinfo {title} {{The mathematical theory of black holes}}}},\ Oxford
  classic texts in the physical sciences\ (\bibinfo  {publisher} {Oxford Univ.
  Press},\ \bibinfo {address} {Oxford},\ \bibinfo {year} {2002})\BibitemShut
  {NoStop}%
\bibitem [{\citenamefont {Geroch}\ \emph {et~al.}(1973)\citenamefont {Geroch},
  \citenamefont {Held},\ and\ \citenamefont {Penrose}}]{GHP_paper}%
  \BibitemOpen
  \bibfield  {author} {\bibinfo {author} {\bibfnamefont {R.}~\bibnamefont
  {Geroch}}, \bibinfo {author} {\bibfnamefont {A.}~\bibnamefont {Held}}, \ and\
  \bibinfo {author} {\bibfnamefont {R.}~\bibnamefont {Penrose}},\ }\href
  {\doibase 10.1063/1.1666410} {\bibfield  {journal} {\bibinfo  {journal}
  {Journal of Mathematical Physics}\ }\textbf {\bibinfo {volume} {14}},\
  \bibinfo {pages} {874} (\bibinfo {year} {1973})},\ \Eprint
  {http://arxiv.org/abs/https://doi.org/10.1063/1.1666410}
  {https://doi.org/10.1063/1.1666410} \BibitemShut {NoStop}%
\bibitem [{\citenamefont {Gustafsson}\ \emph {et~al.}(1995)\citenamefont
  {Gustafsson}, \citenamefont {Kreiss},\ and\ \citenamefont
  {Oliger}}]{gustafsson1995time}%
  \BibitemOpen
  \bibfield  {author} {\bibinfo {author} {\bibfnamefont {B.}~\bibnamefont
  {Gustafsson}}, \bibinfo {author} {\bibfnamefont {H.}~\bibnamefont {Kreiss}},
  \ and\ \bibinfo {author} {\bibfnamefont {J.}~\bibnamefont {Oliger}},\ }\href
  {https://books.google.co.uk/books?id=1JZ2mSQ-O6MC} {\emph {\bibinfo {title}
  {Time Dependent Problems and Difference Methods}}},\ A Wiley-Interscience
  Publication\ (\bibinfo  {publisher} {Wiley},\ \bibinfo {year}
  {1995})\BibitemShut {NoStop}%
\bibitem [{\citenamefont {Brodbeck}\ \emph {et~al.}(1999)\citenamefont
  {Brodbeck}, \citenamefont {Frittelli}, \citenamefont {Hübner},\ and\
  \citenamefont {Reula}}]{doi:10.1063/1.532694}%
  \BibitemOpen
  \bibfield  {author} {\bibinfo {author} {\bibfnamefont {O.}~\bibnamefont
  {Brodbeck}}, \bibinfo {author} {\bibfnamefont {S.}~\bibnamefont {Frittelli}},
  \bibinfo {author} {\bibfnamefont {P.}~\bibnamefont {Hübner}}, \ and\
  \bibinfo {author} {\bibfnamefont {O.~A.}\ \bibnamefont {Reula}},\ }\href
  {\doibase 10.1063/1.532694} {\bibfield  {journal} {\bibinfo  {journal}
  {Journal of Mathematical Physics}\ }\textbf {\bibinfo {volume} {40}},\
  \bibinfo {pages} {909} (\bibinfo {year} {1999})},\ \Eprint
  {http://arxiv.org/abs/https://doi.org/10.1063/1.532694}
  {https://doi.org/10.1063/1.532694} \BibitemShut {NoStop}%
\bibitem [{\citenamefont {Gundlach}\ \emph {et~al.}(2005)\citenamefont
  {Gundlach}, \citenamefont {Martin-Garcia}, \citenamefont {Calabrese},\ and\
  \citenamefont {Hinder}}]{Gundlach:2005eh}%
  \BibitemOpen
  \bibfield  {author} {\bibinfo {author} {\bibfnamefont {C.}~\bibnamefont
  {Gundlach}}, \bibinfo {author} {\bibfnamefont {J.~M.}\ \bibnamefont
  {Martin-Garcia}}, \bibinfo {author} {\bibfnamefont {G.}~\bibnamefont
  {Calabrese}}, \ and\ \bibinfo {author} {\bibfnamefont {I.}~\bibnamefont
  {Hinder}},\ }\href {\doibase 10.1088/0264-9381/22/17/025} {\bibfield
  {journal} {\bibinfo  {journal} {Class. Quant. Grav.}\ }\textbf {\bibinfo
  {volume} {22}},\ \bibinfo {pages} {3767} (\bibinfo {year} {2005})},\ \Eprint
  {http://arxiv.org/abs/gr-qc/0504114} {arXiv:gr-qc/0504114} \BibitemShut
  {NoStop}%
\bibitem [{Note4()}]{Note4}%
  \BibitemOpen
  \bibinfo {note} {For example, this is done in \cite {Hilditch:2019maz}. In
  their case it is more straightforward to do this as with their tetrad
  condition the $\delta $ derivatives only involve ``angular'' derivatives, so
  the Newman-Penrose equations \protect \textup {\hbox {\mathsurround \z@
  \protect \normalfont (\ignorespaces \ref {eq:riem-11}\unskip \@@italiccorr
  )}}-\protect \textup {\hbox {\mathsurround \z@ \protect \normalfont
  (\ignorespaces \ref {eq:riem-13}\unskip \@@italiccorr )}} provide a
  convenient set of constraint equation at the $r=r_{min}$}\BibitemShut
  {NoStop}%
\end{thebibliography}%

\end{document}